\documentclass[preprint,showpacs,preprintnumbers,amsmath,amssymb]{revtex4}
\usepackage[dvips]{graphicx}
\usepackage{amsmath}

\begin{document}

\title{How the asymmetry of internal potential influences the shape of
$I-V$ characteristic of nanochannels}

\author{I. D. Kosi\'nska} \email[e-mail address:~]{kosinska@th.if.uj.edu.pl}

\address{
M. Smoluchowski Institute of Physics, Jagiellonian University,
Reymonta 4, PL-30-059 Krak\'ow, Poland\\
}

\begin{abstract}

Ion transport in biological and synthetic nanochannels is
characterized by such phenomena as ion current fluctuations,
rectification, and pumping. Recently, it has been shown that the
nanofabricated synthetic pores could be considered as analogous to
biological channels with respect to their transport
characteristics \cite{Apel, Siwy}. The ion current rectification
is analyzed. Ion transport through cylindrical nanopores is
described by the Smoluchowski equation. The model is considering
the symmetric nanopore with asymmetric charge distribution. In
this model, the current rectification in asymmetrically charged
nanochannels shows a diode-like shape of $I-V$ characteristic. It
is shown that this feature may be induced by the coupling between
the degree of asymmetry and the depth of internal electric
potential well. The role of concentration gradient is discussed.

\end{abstract}

\maketitle

\section{Introduction}

The role of ion transport through narrow protein channels for
living cells has been widely studied. There are many experimental
and theoretical attempts in understanding the activity of
biological channels at physiological conditions. Among models of
cellular electrical activity the Goldman-Hodgkin-Katz (GHK)
current equation plays an important role. It is worth noting that
the equation assumes of a constant electric field. There are other
models in biophysical literature e.g. barrier models with several
free-energy barriers within the channel \cite{Eyring, Woodbury,
Lauger}. These models are phenomenological, i.e. the potential
energy barriers used in the models often do not correspond to
physical properties of the channel. Although they can give good
quantitative descriptions of some experimental data \cite{Sneyd},
they fail to fit these data sets which include $I-V$ relations
measured in asymmetrical solutions \cite{Nonner}. One of the
reasons is that their the electrostatics is modeled incorrectly.
On the other hand, the Poisson-Nernst-Planck (PNP) model, based on
the mean field approximation, is an alternative theory. Apart from
its own limitations \cite{Lewitt,Corry2,Corry}, the 3DPNP model
seems to take into account electrostatics correctly \cite{Nonner}.
However, in general, it is not possible to obtain an exact
solution to the Nernst-Planck (NP) equation, coupled to the
electric field by means of the Poisson equation. That is why we
consider a simplified system that allows for intuitive
understanding of the underlying mechanism of examined phenomenon.

The intention of the paper is to demonstrate that the asymmetry of
the internal physical field also plays a significant role in
permeation. To prove this conjecture we focus on the only one
source of the asymmetry, viz. the electrostatics. Thus,
geometrical effects of the channel are purposely omitted by choice
of the cylindrical structure of the channel. Chemical structure is
not considered, either. Factors such as geometry, chemical
structure, etc. may also be sources of the potential asymmetry.
The selectivity filters seem to be governed by different
mechanisms \cite{Laio}, therefore they are not discussed here. We
propose a simplified continuous model that is able to explain the
experimental behavior of interest.

The studies of the synthetic channels, which in several aspects
resemble the biological ones \cite{Siwy2, Siwy4}, may be helpful
in investigating the mechanism of ionic transport. The ionic
currents through these nanochannels exhibit several peculiarities.
One of them is the asymmetric and non-linear shape of
current-voltage characteristic both for different biological
channels \cite{{kienker},{okazaki},{Gor}} and for asymmetrically
shaped synthetic ones \cite{{Apel}, {Siwy}, {Siwy3}, {Ful1}}.
These asymmetries in the I-V characteristics are related, among
others, to the nanopumping mechanism \cite{SiF} and to the
asymmetry of nanodiffusion \cite{Kos1, Kos2}. It is worth
mentioning that the role of spatial/time asymmetry in large number
of physical processes was discussed in \cite{Hanggi}, and that the
experiment has shown that rectification occurs for colloidal
particles in a microfabricated channel with a topological
ratchet-like polarity \cite{Marquet}. The above-mentioned results
point out that the theoretical explanation of current
rectification in nanochannels should enclose the intrinsic
asymmetry of channels.

The aim of this work is to stress the role of the asymmetry of
potential well inside the channel in the rectification process and
in the diode-like shape of $I-V$ dependence. In our model the
total current $I$ through the channel is driven both by an
external and an internal electrical field. We shall investigate
the shape of $I-V$ curve for symmetric and asymmetric cases of the
potential well and for different depths of the well. For this
purpose we need the given shape of the potential well inside the
channel. Thus, the first problem we shall consider is that of
determining the electric potential $\phi_{int}$ inside the channel
due to prescribed surface density $\sigma$. For simplicity, we
shall consider a coaxial infinitely long cylindrical channel and
we take the radius $R$ of the cylinder small enough to allow the
ions go only through the channel along $z$-axis, which implies the
lack of electrolytic solution inside the channel (no screening).
If we assume the equilibration in the transverse direction of the
channel \cite{Mon} we can operate with the $z$-dependent potential
$\phi^{r=0}_{int}(z)$ on the $z$-axis only. Therefore our problem
is reduced from the three-dimensional to the one-dimensional
description. We shall discuss the evolution of the probability
density of the finding an ion inside the channel, therefore we
shall use the 1-D Smoluchowski equation.

The Smoluchowski equation gives the conditional probability that
the particle starting from the point $z(t_0)$ reaches the point
$z$ at the time $t$ \cite{Smol, Ful}. It describes the {\it
diffusion of probability}, because the process of diffusion is the
superposition of Brownian motions of the molecules of the
substance under consideration \cite{Smol, Smol2}. In an open
state, the measurable electric current through the biological
channels is at the picoampere level which gives about $10^8$ ions
per second \cite{Kuyucak}. Thus one ion passes the channel at the
tens of nanosecond and after $10^8$ passages we get the measurable
electric current. These estimations enable us to pass from one-ion
description via probability density, i.e. from the Smoluchowski
equation to the continuous description in terms of the electric
current density, i.e. to the Smoluchowski-Nernst-Planck equation.

\section{Model}

The problem which we want to consider is that of electric current
flowing through cylindrical channel. There is a broad collection
of papers in which the cylindrical synthetic nanotubes \cite{{1},
{NJP},{Schattat}, Ful1} and biological cylindrical ionic channels,
e.g. gramicidin channel
\cite{Woolf}, Class 1 porins
\cite{porin} are discussed. Let us consider a dielectric membrane
(biological or synthetic) separating two large regions of space
that contain some electrolytic solutions. There is a pore in the
membrane (one-dimensional channel of a radius $R$ and a length
$L$, see Fig. 1) through which the ions can move more or less
freely.

\begin{figure}[htb!]
    \begin{center}
      \includegraphics*[width=3.25in]{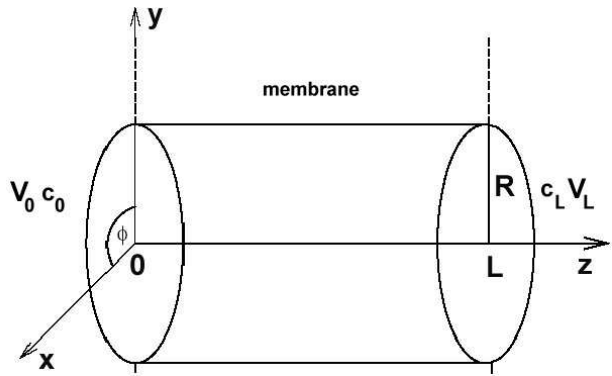}
       \caption{\small{Channel in cartesian coordinates ($x,y,z$) with the $z$
axis along the axis of the cylinder; the cylinder has a radius $R$
and a length $L$; $V_0, V_L$ are potentials and $c_0,c_L$
concentrations on the left and on the right side of the channel,
respectively.}}
    \end{center}
    \label{fig:1}
\end{figure}

The biological ion channels are ion selective \cite{Hille}. The
same feature is displayed in some synthetic channels \cite{Siwy,
Lev}. Therefore, for simplicity, we assume that only positive ions
can enter our pore. In Cole's model \cite{Cole} of electrical
activity in membranes the driving force for the ions is given by
the difference between the membrane potential
$U=V_{in}(0)-V_{out}(L)$ and the reversal potential, which is the
potential at which the current is zero. In the
Goldman-Hodgkin-Katz model (GHK model) of ion flux through the
channel the simplifying approximation is made that the potential
gradient through the channel is constant \cite{Fall}. On the other
hand in experiments with synthetic membranes the driving force is
caused by electrodes of potential $V_+$ and $V_-$, located in the
electrolyte at macroscopic distances from the membrane (Fig. 1 in
\cite{Ful1}). Thus in both cases we put the related potential in
the form:
\begin{equation}
\phi_{ext}(z) = V_+ - Uz/L,
\end{equation}
where $U=V_+-V_-$ and $L$ is a length of the channel. (In general,
ions moving through the channel affect the local electric field.
Thus, the related electric potential may be given by a more
complicated function. However, if the channel is short or the
ionic concentrations on either side of the membrane are small this
approximation seems to be correct \cite{Sneyd}.)

We consider charged channel with a given shape of internal
potential $\phi^{r=0}_{int}(z)$ along the $z$ axis. Therefore, the
total electric field $\vec{E}$ acting on ions inside the channel
is a linear superposition of the external $\vec{E}_{ext}$ and the
internal $\vec{E}_{int}$ fields, the latter one induced by charges
on the wall of the channel. The goal of the following analysis is
to demonstrate that the field asymmetry is sufficient to observe
diode-like shape of $I-V$ relation. Let us consider two shapes of
the internal potential $\phi^{r=0}_{int}(z)$ described below and
presented in Fig.2. They roughly correspond to two types of
channels, short biological (Fig. 2a) and long synthetic ones (Fig.
2b).

\begin{figure}[htb!]
    \begin{center}
      \includegraphics*[width=2.25in]{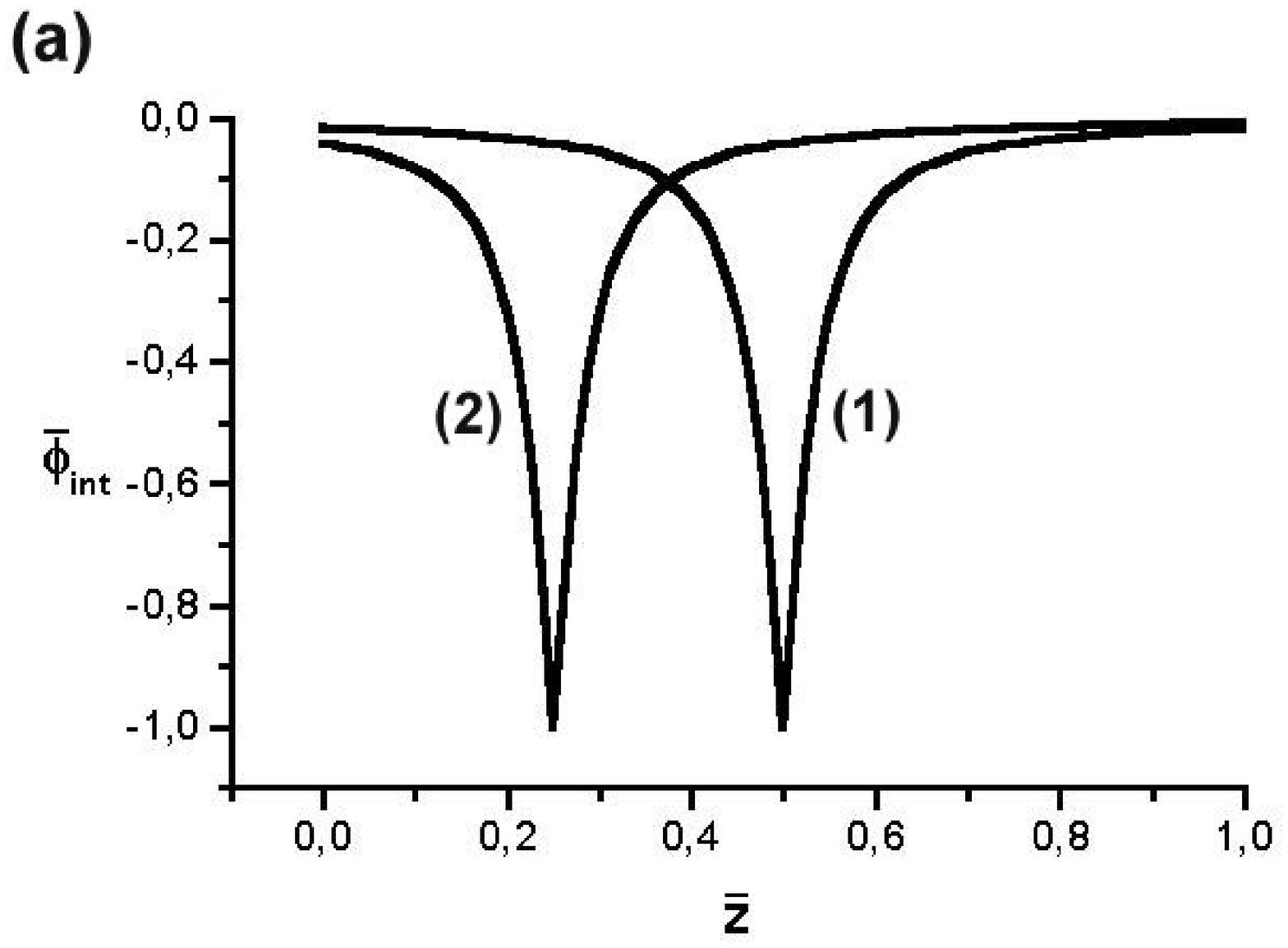}
      \includegraphics*[width=2.25in]{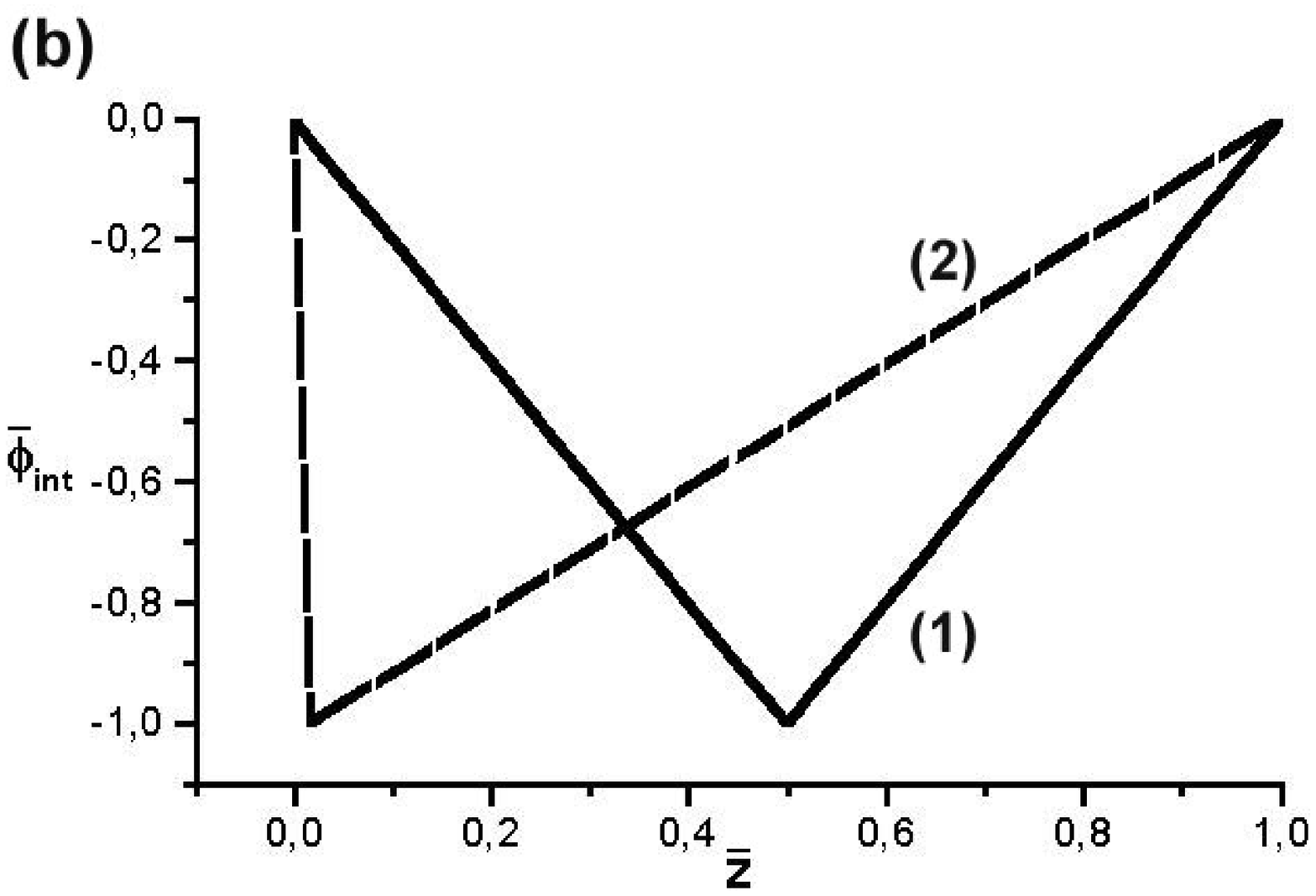}
       \caption{\small{The shape of potential well $\overline{\phi}_{int}$ (we
use dimensionless coordinate $\overline{z}=z/L$ and potential
$\overline{\phi}_{int}=\beta e \phi_{int}$ a) for
$\epsilon_{out}=2\epsilon_0$, $\epsilon_{int}=80\epsilon_0$,
$\sigma_g= 0,015$ e/nm$^2$, $\overline{R}=0.004$ and
$\overline{g}=0.5$ ((1)-line) and $\overline{g}=0.25$ ((2)-line);
b) for ratchet-like shape with
$\overline{\phi}(\overline{L}_1=0)=\overline{\phi}(\overline{L}_2=1)=0$,
$\overline{\phi}(\overline{g})=-1$ and $\overline{g}=0.5$
((1)-line) and $\overline{g}=0.008$ ((2)-line).}}
    \end{center}
    \label{fig:2}
\end{figure}

The first potential shape we propose to consider is the potential
$\phi_{int}$ of the form shown in Fig. 2a. This shape of the
potential well, which results from charged residues localized at
$z=g$, might simulate the situation in short biological [e.g.
\cite{porin}] and synthetic \cite{Schattat} channels. To find this
shape we use the Eq.~\ref{eqn:app} (see Appendix) and we put the
surface charge density into the formula (\ref{eqn:pow}) to be
$\sigma(z)=\sigma_g\delta (z-g)$. The value of the parameter $g$
gives us the asymmetry of the function $\phi_{int}$.

The second shape, which corresponds to continuous charge
distribution, is the electrostatic potential $\phi_{int}$ in the
``ratchet-like'' shape (Fig. 2b). This case seems to be more
adequate to the long synthetic tubes in which the ratio of length
to width goes to 0, thus the approximation of an infinitely long
cylinder works well \cite{Ful1}.
\begin{equation}
\phi^{r=0}_{int}(z)= \left\{
\begin{array}{ccc}
\vspace{10pt} \phi(g)\Delta\phi' (z-L_1)&~~\Delta \phi' =
\displaystyle\frac{1-\phi(L_1)/\phi(g)}{g-L_1},&~
\rm{for}~~z\in [L_1,g]\\
\vspace{10pt} \phi(g)\Delta \phi'' (z-L_2)&~~\Delta \phi'' =
\displaystyle\frac{1-\phi(L_2)/\phi(g)}{g-L_2},&~\rm{for}~~z\in
(g,L_2]\\
0&&\rm{elsewhere}
\end{array}
\right.\label{eqn:ratchet_eq}
\end{equation}
where $g$ is the asymmetry parameter of this function, $\phi(g)$
is the value of the potential for $z=g$. The asymmetry of the
potential is controlled by the value of the parameter $g$ (for
$g=(L_1+L_2)/2$ and $L_1=L-L_2$ we have the symmetric function,
the most asymmetric case we have if $g\to L_1$ (or $g\to L_2$)).

\section{Results}

The kinetic Smoluchowski equation is commonly used in various
physical, chemical, etc., problems \cite{Ful}. The equation
describes diffusion in any physical field. Therefore a model based
on this equation can be placed between these two GHK and PNP
models (the GHK equation is the 1D Smoluchowski equation in a
constant field).

Let us start from the 1D Smoluchowski equation that contain the
electrostatic field:
\begin{equation}
\begin{array}{c}\vspace{10pt}
\displaystyle\frac{\partial}{\partial t}p(z,t) = - \frac{\partial}{\partial z}j(z,t),\\
\displaystyle j(z,t) = -D\frac{\partial}{\partial z} p(z,t) + \mu
E_z p(z,t) \label{eqn:smol}
\end{array}
\end{equation}
where $E_z =-\displaystyle\frac{\partial}{\partial z} \phi(z)$ is
the $z$-component of electric field, $j(z,t)$ denotes probability
density current, which describes the flux produced by diffusion of
cations. They are driven by the difference in their probabilities
of entering the pore from the left $p(0)$ and from the right
$p(L)$, respectively and by ionic migration caused by difference
in electric potential. We assume here that the mobility of a
particle $u$ fulfills the Nernst-Einstein equation, the diffusion
coefficient $D=u/\beta$, where $\beta = 1/k_BT$. The ion mobility
$\mu=ez_cu$, where $e$ is the electric charge, and $z_c$ is the
cation valence.

We can write the Eq.~\ref{eqn:smol} in more convenient form:
\begin{equation}
\displaystyle\frac{\partial p(z,t)}{\partial t} = \frac{
\partial}{\partial z} D\displaystyle e^{- z_c \beta e\phi(z)} \frac{
\partial}{\partial z}
e^{ z_c \beta e \phi(z)} p(z,t)=-\frac{
\partial}{\partial z} j(z,t).
\end{equation}

The measurable quantity is the stationary mass current $J_i$
($J_i\ne J_i(t)$) flowing through the channel cross-section
$\mathcal{A}=\pi R^2$:
\begin{equation}
J_i=\pi R^2j(z) \label{eqn:nat_mas}
\end{equation}
(we use here the identification mentioned in the introduction i.e.
one ion passes the channel at the tens of nanosecond and after
$10^8$ passages we get the measurable electric current).

For simplicity we use dimensionless coordinate $\overline{z}=z/L$,
and potential $\overline{\phi}=\beta e\phi$. For the potassium
cations $z_c=1$. Thus, the electric current resulting from mass
current is:
\begin{equation}
I =FJ= F\displaystyle\pi R^{2}\frac{D}{L}
\displaystyle\frac{c_0e^{\overline{\phi}_{int}(0)}-c_1e^{\overline{\phi}_{int}(1)-
\overline{U}}}{\displaystyle\int^1_0 d\overline{z}
e^{\overline{\phi}_{int}(\overline{z})-\overline{U}\overline{z}}},\label{eqn:prad}
\end{equation}
where $F$ is the Faraday's constant. If we put
$\overline{\phi}_{int}=0$ we obtain the famous
Goldman-Hodgkin-Katz (GHK) current equation.

In the case of ratchet-like shape of potential
$\overline{\phi}_{int}$ Eq.~\ref{eqn:ratchet_eq} with
$\overline{L}_1=0$ and $\overline{L}_2=1$ we get the analytic
solution:
\begin{equation}
\begin{array}{c}
\vspace{10pt}
I =FJ= \\
F\displaystyle\pi R^{2}\frac{D}{L}
(c_{0}e^{\overline{\phi}(0)}-c_{1}e^{\overline{\phi}(1)-
\overline{U}})/\bigg(\displaystyle\frac{1-e^{\overline{\phi}
(\overline{g})-\overline{\phi}(0)-\overline{U}\overline{g}}}
{\overline{U}-\overline{\phi}(\overline{g})
\Delta\overline{\phi}'}
+\frac{e^{\overline{\phi}(\overline{g})-\overline{\phi}(1)
-\overline{U}\overline{g}}-e^{-\overline{U}}}{\overline{U}-
\overline{\phi}(\overline{g})\Delta\overline{\phi}''}\bigg).\label{eqn:prad_b}
\end{array}
\end{equation}

Now, we analyze $I-V$ relations for two kinds of potential
function $\overline{\phi}_{int}(z)$ that are described in {\it
Sec. Model}. Let us start from the second case being ratchet-like
function of $\overline{z}$. We use the analytic solution for
current that is given by Eq.~\ref{eqn:prad_b} where we put
$\overline{\phi}_{int}(\overline{L}_1=0)=\overline{\phi}_{int}(\overline{L}_2=1)=0$,
$\overline{R}=R/L=0.0004$, and $D_{K^{+}}=2*10^{-9}$ m$^2$/s. The
ion current through the nanochannel with no gradient of
concentration ($c_{0}=c_{1}=0.01$ M) depends on both the depth of
the potential well and on the degree of asymmetry of internal
potential $\phi_{int}$ (see Figures 3). For the fixed depth of
potential well $\overline{\phi}(\overline{g})=-9$ and different
values of $\overline{g}$ (from 0.5 to 0.00001) we get various
degrees of rectification. It can be clearly seen that with
increasing asymmetry ($\overline{g}\to 0$) the $I-V$
characteristic tends to a diode-like shape (Fig. 3a). However the
depth of potential well $\overline{\phi}(\overline{g})$ is
important as well. For the lower value of
$\overline{\phi}(\overline{g})=-1$ (Fig. 3b) we observe non-linear
shape of $I-V$ curve for both $\overline{g}=0.5$ and
$\overline{g}=0.00001$. For the asymmetric potential well the
$I-V$ characteristic shows asymmetry. On the other hand, for the
higher value of $\overline{\phi}(\overline{g})=-9$ (Fig. 3c) the
difference between symmetric and asymmetric case is much stronger.
In Fig. 3c the $I-V$ curve clearly shows the rectification
(diode-like shape). The direction of the rectification depends on
the value of $\overline{g}$ i.e. if it changes from 0.5 to 0.00001
or from 0.5 to 0.99999.

\begin{figure}[htb!]
    \begin{center}
      \includegraphics*[width=2.25in]{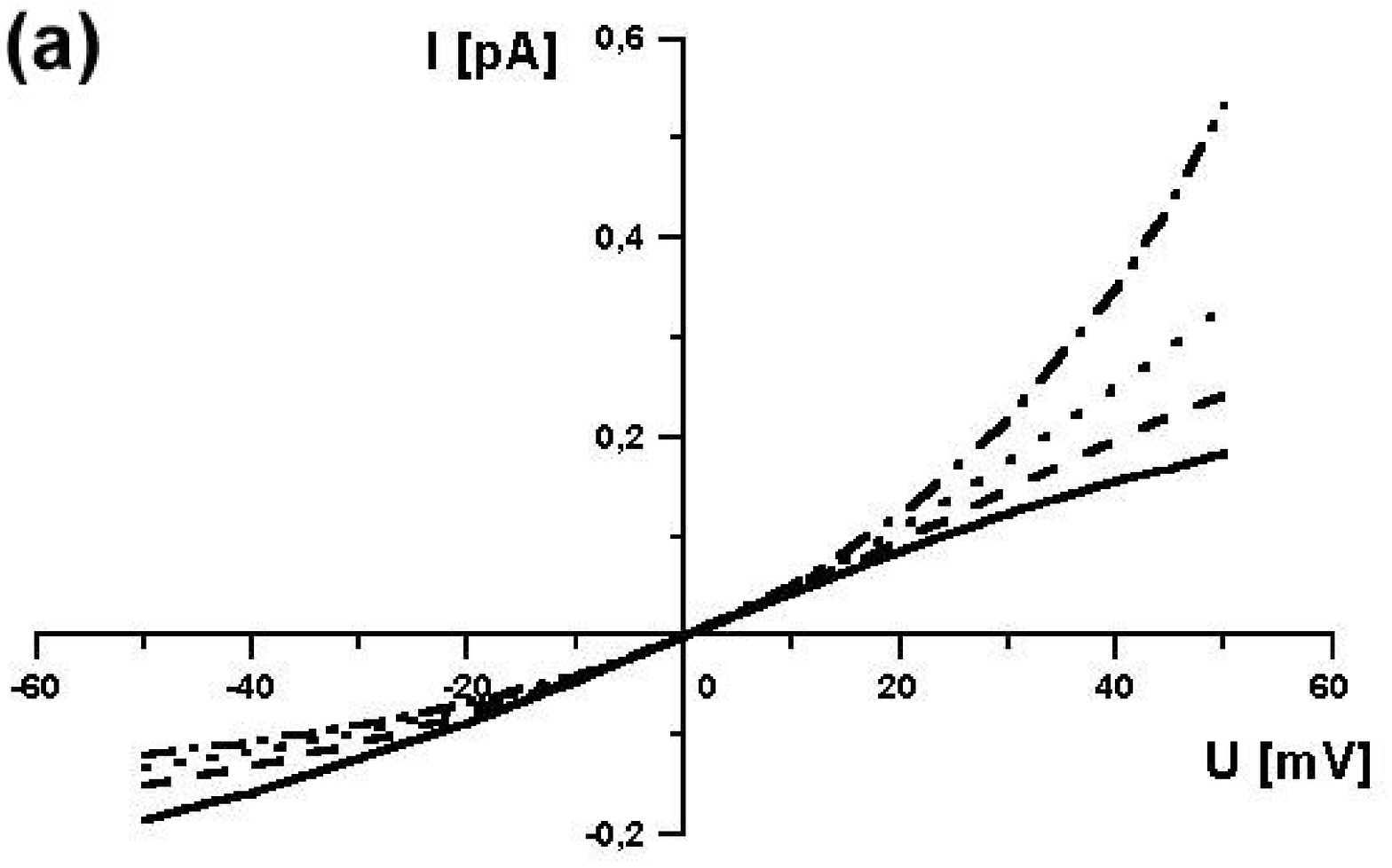}
      \includegraphics*[width=2.25in]{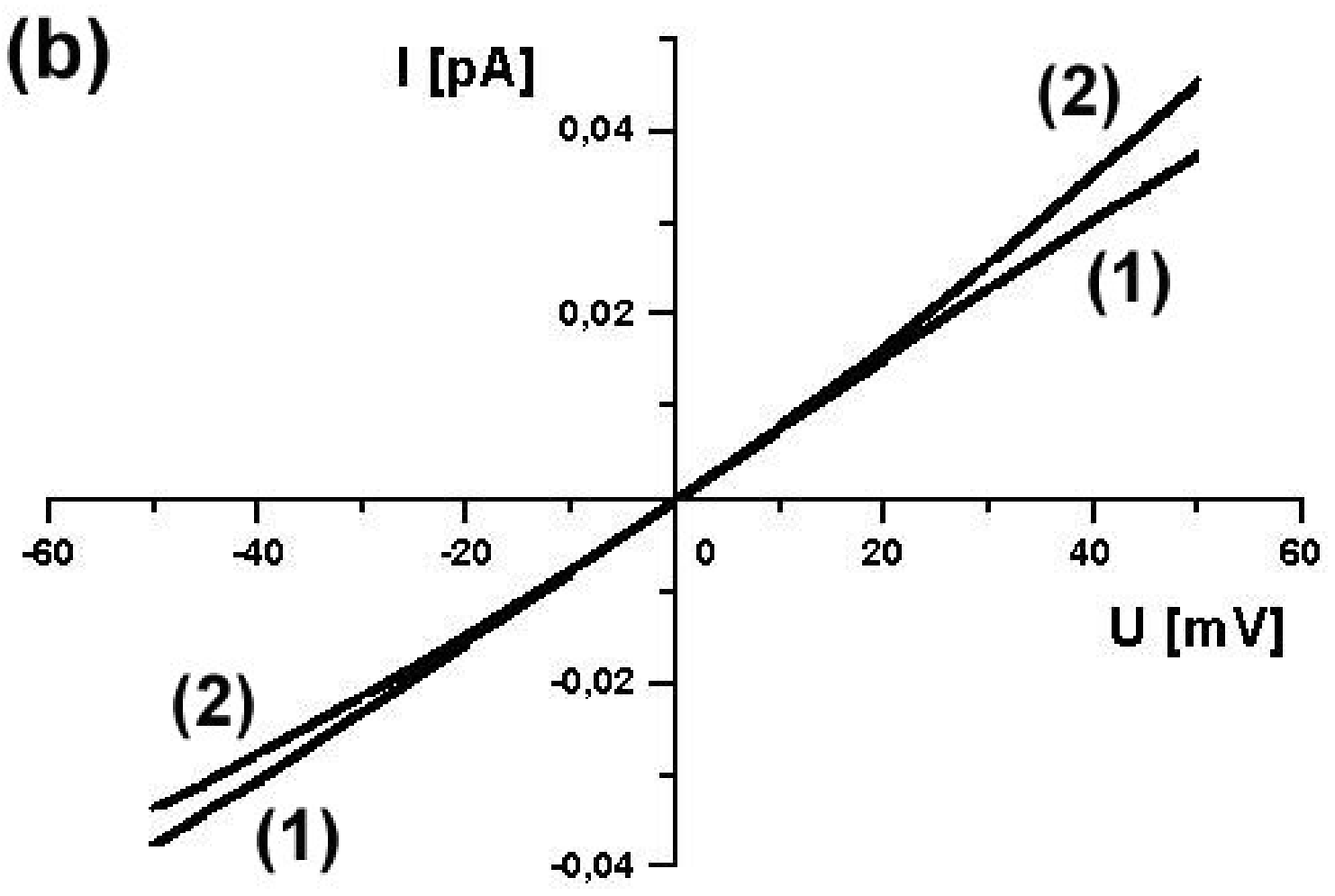}
      \includegraphics*[width=2.25in]{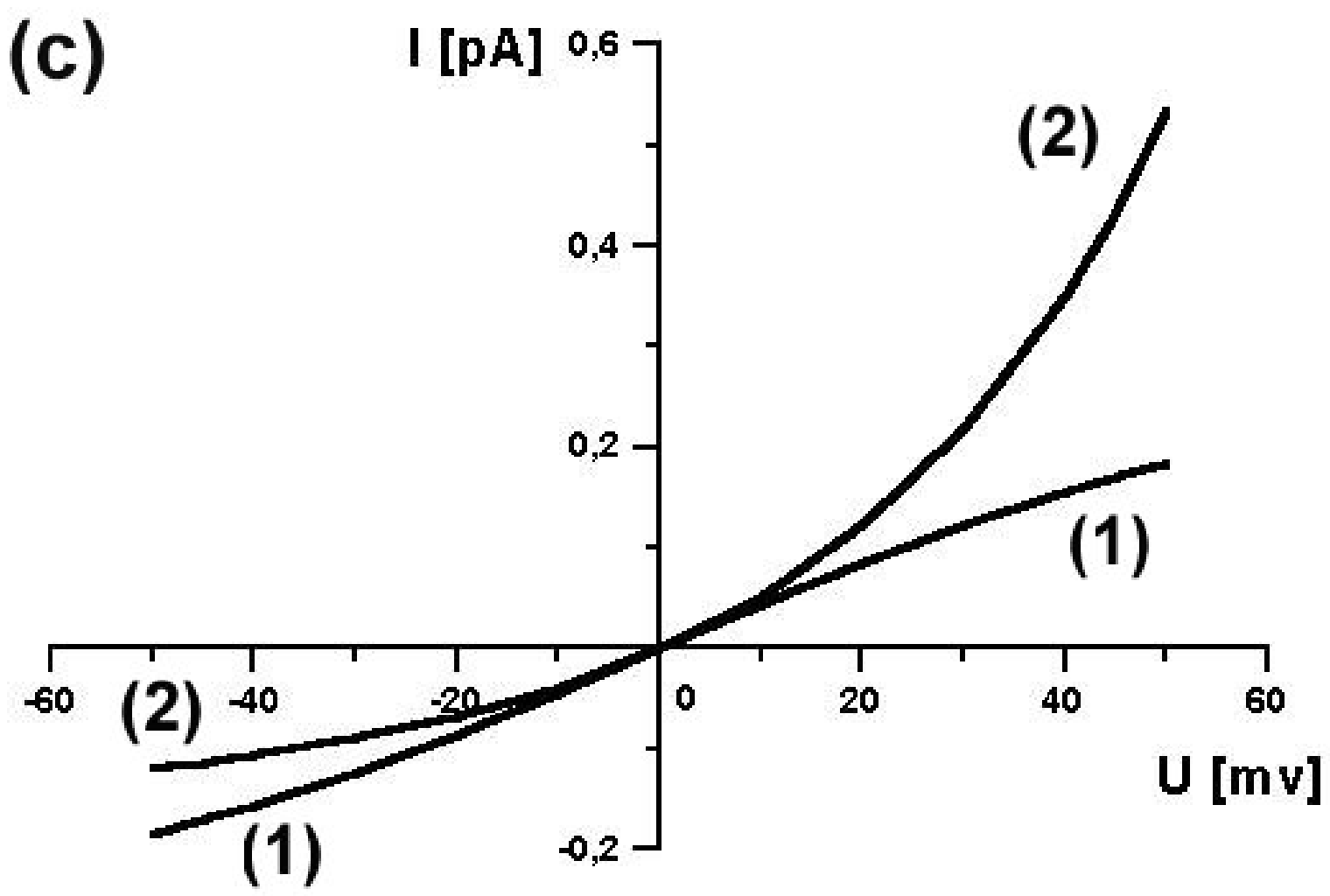}
       \caption{\small{$I-V$ dependence for the $\overline{\phi}_{int}$ being
ratchet-like function of $\overline{z}$, for
$\overline{\phi}_{int}(\overline{L}_1=0)=\overline{\phi}_{int}(\overline{L}_2=1)=0$,
$\overline{R}=R/L=0.0004$, $c_{0}=c_{1}=0.01$ M,
$D_{K^{+}}=2*10^{-9}$ m$^2$/s; a)
$\overline{\phi}(\overline{g})=-9$ and $\overline{g}=0.5$ - solid
line, $\overline{g}=0.3$ - dashed line, $\overline{g}=0.15$ -
dotted line, $\overline{g}=0.00001$ - dot-dashed line; and b)
$\overline{\phi}(\overline{g})=-1$, c)
$\overline{\phi}(\overline{g})=-9$; in both cases for
$\overline{g}=0.5$ - symmetric shape of potential function
((1)-line) and for $\overline{g}=0.00001$ - the asymmetric shape
((2)-line).}}
    \end{center}
    \label{fig:3}
\end{figure}

For the potential $\overline{\phi}_{int}$ showed in Fig. 2a (where
$\overline{R}=R/L=0.004$), the values of boundary condition
$\overline{\phi}(0)$ and $\overline{\phi}(1)$ differ from 0. What
is more in the asymmetric case ((2)-line in Fig.4) the difference
between them grows with increasing value of the potential well
depth $\overline{\phi}(\overline{g})$ i.e. for
$\overline{\phi}(\overline{g})=-10$, $\overline{\phi}(0)=-0.608$
and $\overline{\phi}(1)=-0.203$ whereas for
$\overline{\phi}(\overline{g})=-40$, $\overline{\phi}(0)=-2.430$
and $\overline{\phi}(1)=-0.811$. For $c_0=c_1$ and symmetric
potential function $\overline{\phi}_{int}$ the $I-V$ dependence is
symmetric and weakly non-linear in both cases of
$\overline{\phi}(\overline{g})$ ((1)-line in Fig. 4). For the
asymmetric potential $\overline{\phi}_{int}$, we observe a weak
non-linearity and the appearance of the reversal potential for the
channel (Fig. 4a) which for increasing value of
$\overline{\phi}(\overline{g})$ grows up (Fig. 4b). Therefore in
that case we do not obtain the diode-like shape for the $I-V$
curve. Note that these results reproduce the data reported in
\cite{Kos1}.

\begin{figure}[htb!]
    \begin{center}
      \includegraphics*[width=2.25in]{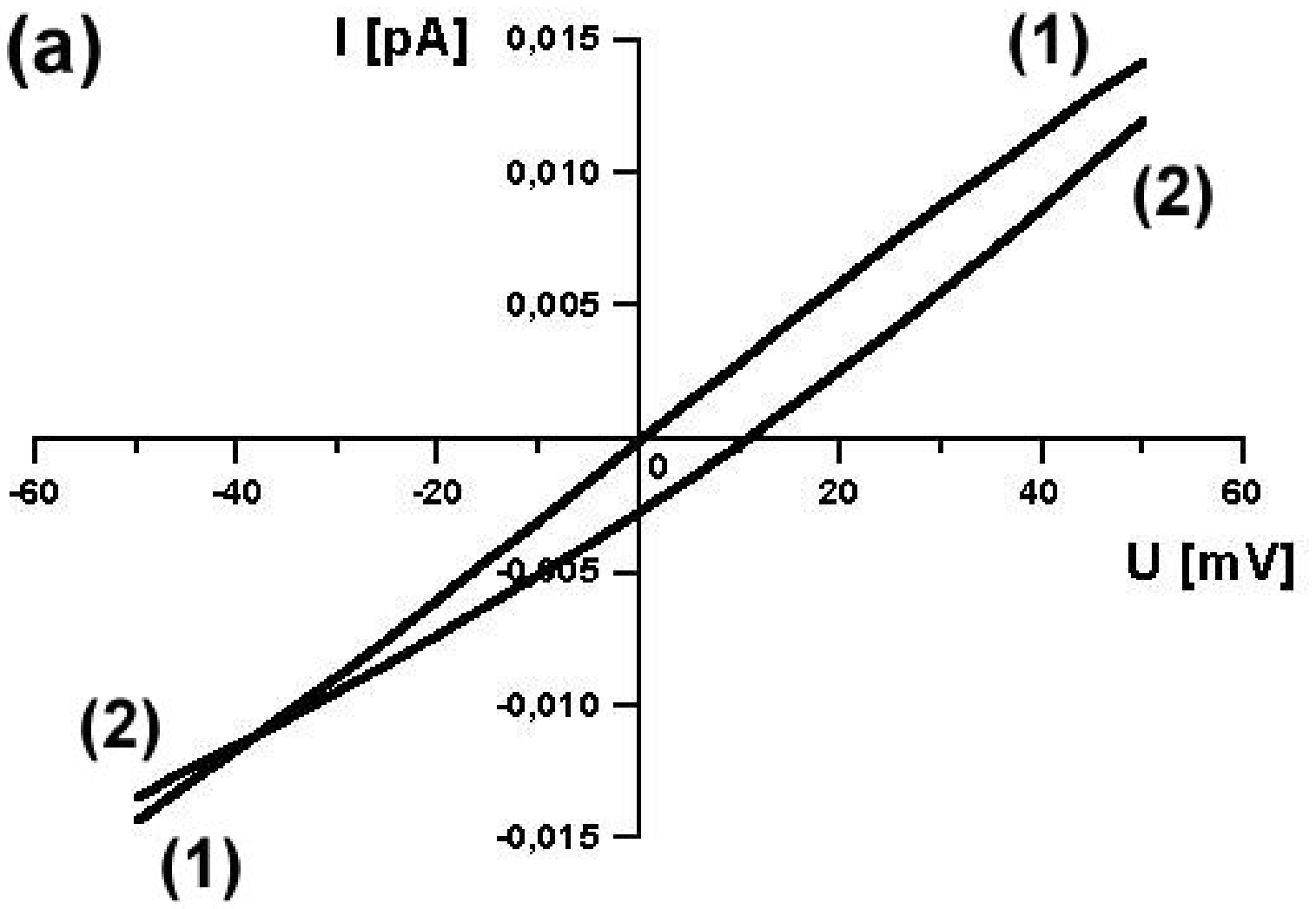}
      \includegraphics*[width=2.25in]{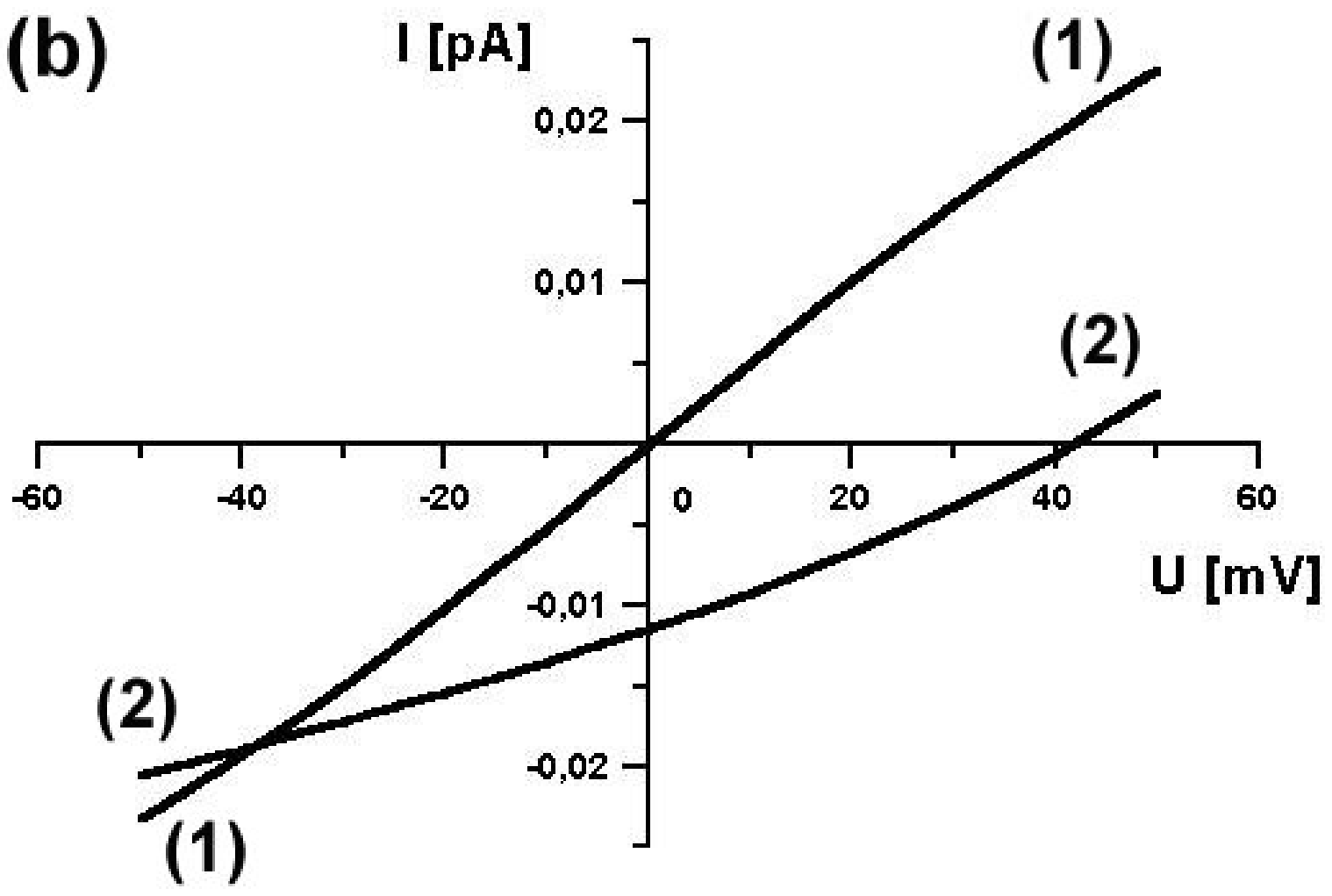}
       \caption{\small{$I-V$ curve for the $\overline{\phi}_{int}$ being
localized function of $\overline{z}$, for
$\overline{R}=R/L=0.004$, $c_{0}=c_{1}=c=0.01$ M,
$D_{K^{+}}=2*10^{-9}$ m$^2$/s, the depth of potential well and the
boundary conditions: a) $\overline{\phi}(\overline{g})=-10$,
$\overline{\phi}(0)= \overline{\phi}(1)=-0.302$ for (1)-line,
$\overline{\phi}(0)=-0.608$, $\overline{\phi}(1)=-0.203$ for
(2)-line; b) $\overline{\phi}(\overline{g})=-40$,
$\overline{\phi}(0)=\overline{\phi}(1)=-1.208$ for (1)-line,
$\overline{\phi}(0)=-2.430$, $\overline{\phi}(1)=-0.811$ for
(2)-line; in both cases for $\overline{g}=0.5$ - symmetric shape
of potential function ((1)-line) and for $\overline{g}=0.25$ - the
asymmetric shape ((2)-line).}}
    \end{center}
    \label{fig:4}
\end{figure}

In described-above cases we put $c_0=c_1$. However the boundary
condition $\overline{\phi}(0)\ne\overline{\phi}(1)$ simulates the
gradient of concentration $c_0\ne c_1$ in Eq.~\ref{eqn:prad}. Thus
putting $c_1/c_0=e^{\overline{\phi}(0)}/e^{\overline{\phi}(1)}$
recovers the diode-like shape (Fig. 5).

\begin{figure}[htb!]
    \begin{center}
      \includegraphics*[width=2.25in]{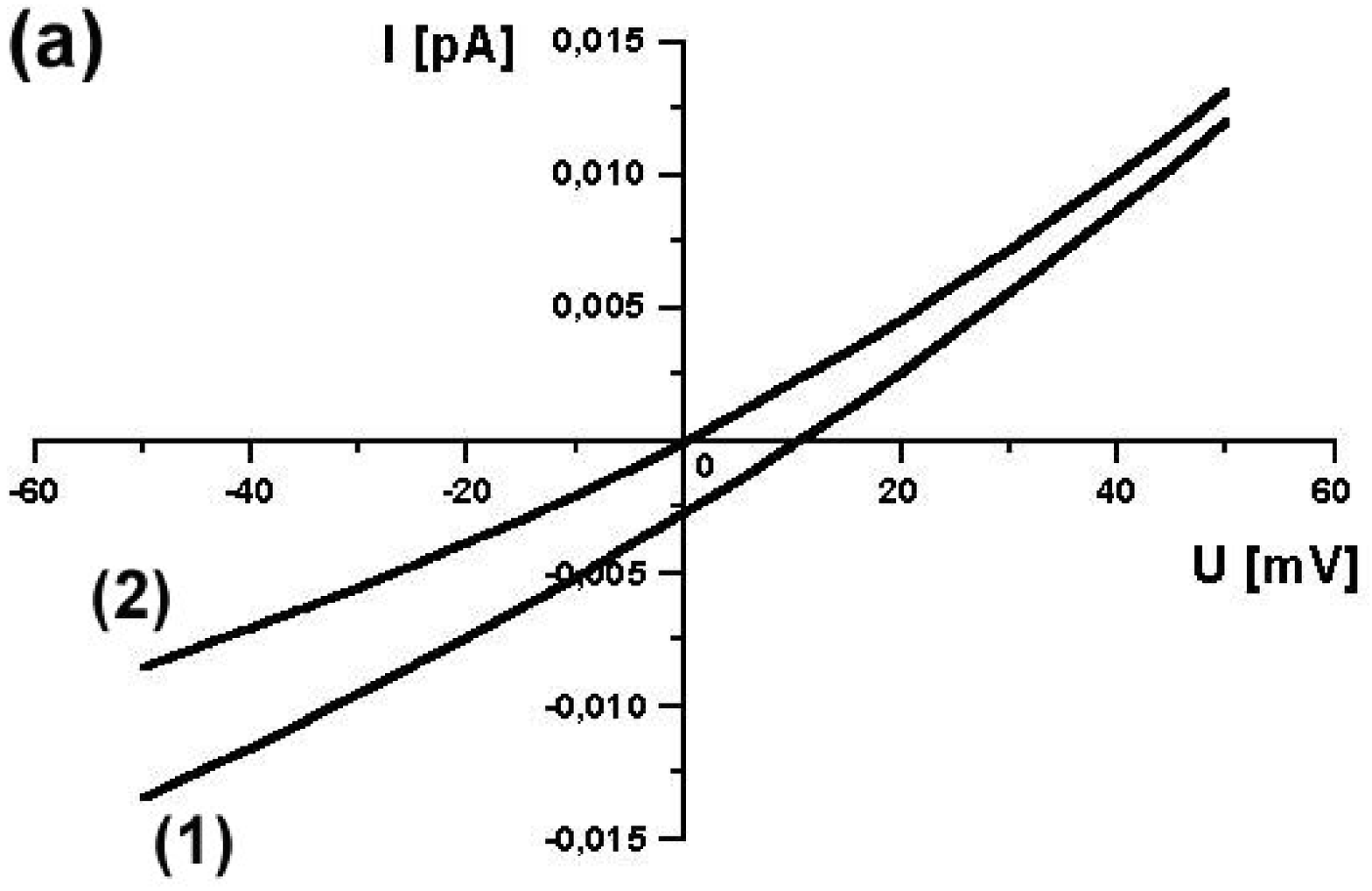}
      \includegraphics*[width=2.25in]{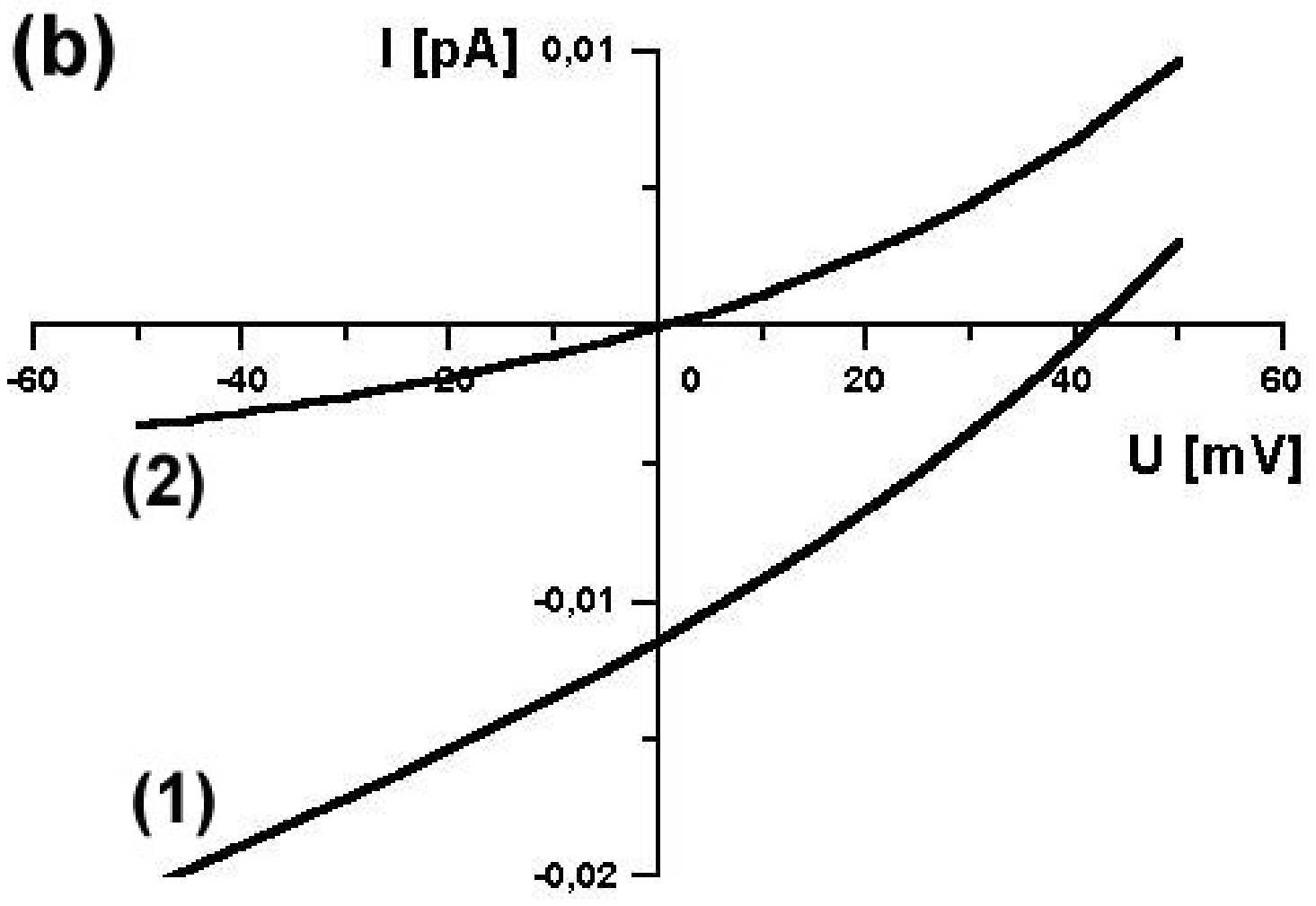}
       \caption{\small{$I-V$ curve for the $\overline{\phi}_{int}$ being
localized function of $\overline{z}$, $\overline{g}=0.25$ - the
asymmetric shape of potential function, $\overline{R}=R/L=0.004$,
$c_{0}=0.01$ M, $D_{K^{+}}=2*10^{-9}$ m$^2$/s, (1)-line for
$c_1=c_0$, (2)-line for
$c_1=c_0e^{\overline{\phi}(0)}/e^{\overline{\phi}(1)}$, the depth
of potential well and boundary condition: a)
$\overline{\phi}(\overline{g})=-10$, $\overline{\phi}(0)=-0.608$,
$\overline{\phi}(1)=-0.203$; b)
$\overline{\phi}(\overline{g})=-40$, $\overline{\phi}(0)=-2.430$,
$\overline{\phi}(1)=-0.811$.}}
    \end{center}
    \label{fig:5}
\end{figure}

One can obtain similar results of $I-V$ dependence for several
other shapes of $\phi_{int}$ with cases of boundary conditions
discussed-above.

\begin{figure}[htb!]
    \begin{center}
      \includegraphics*[width=2.25in]{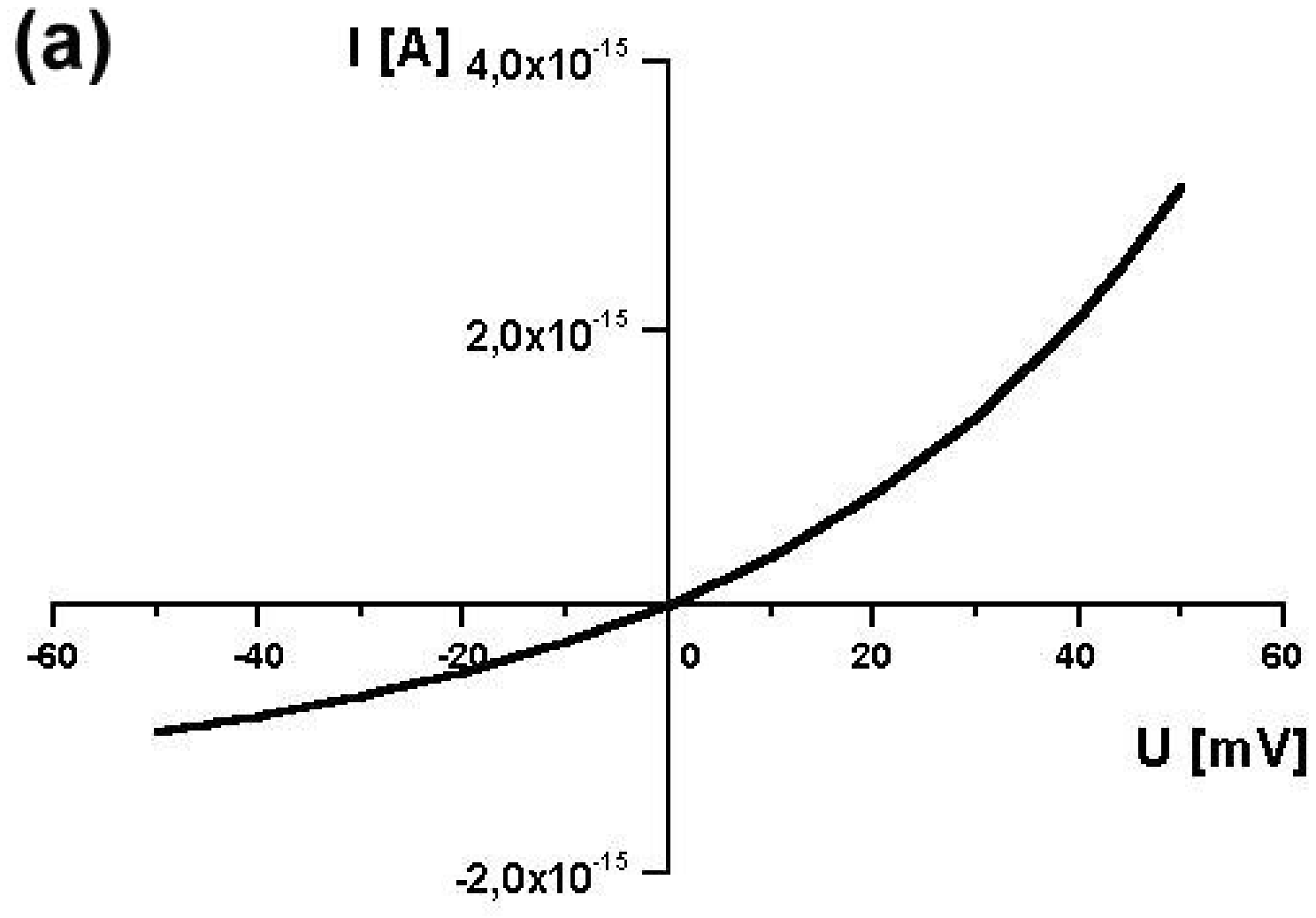}
      \includegraphics*[width=2.25in]{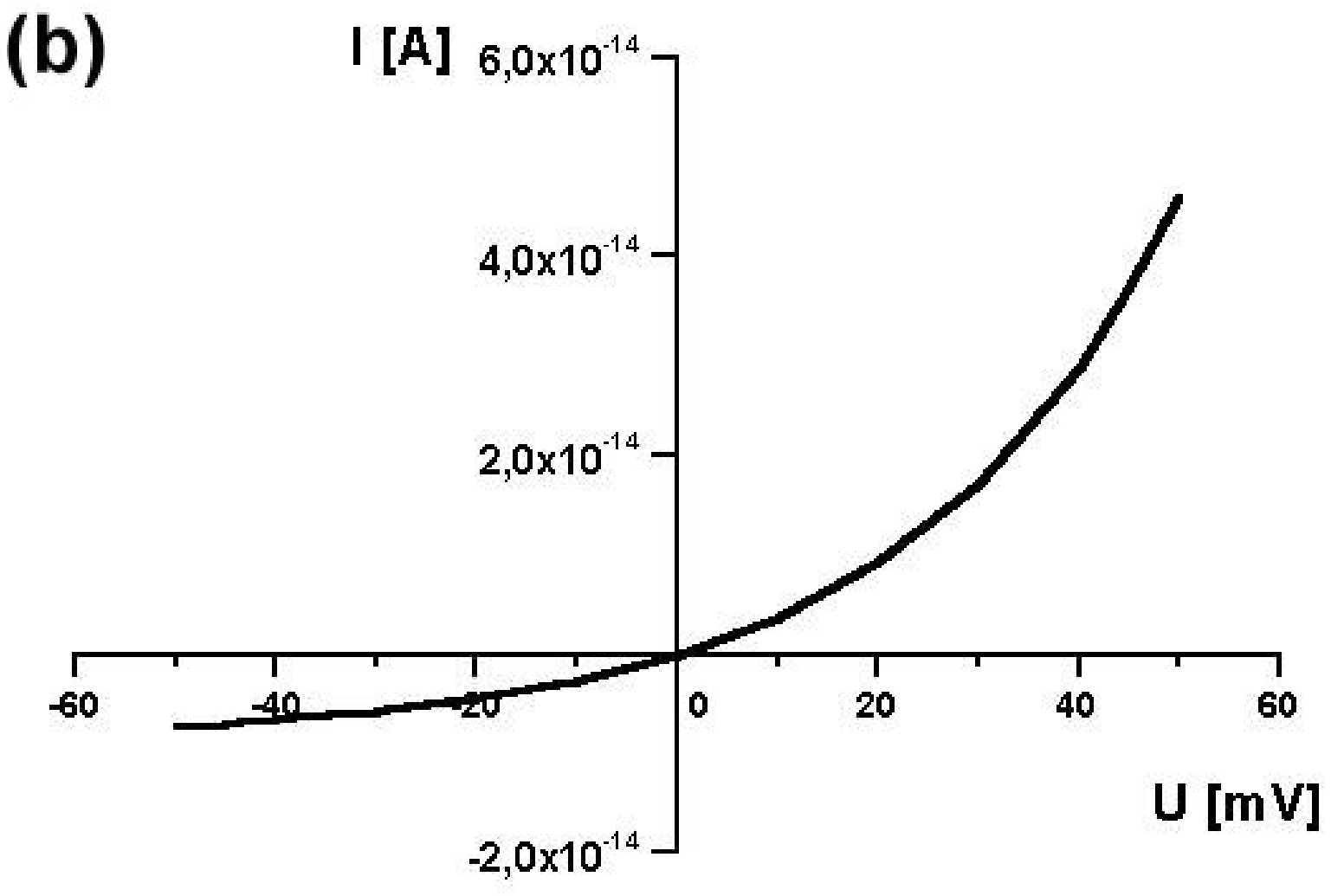}
       \includegraphics*[width=2.25in]{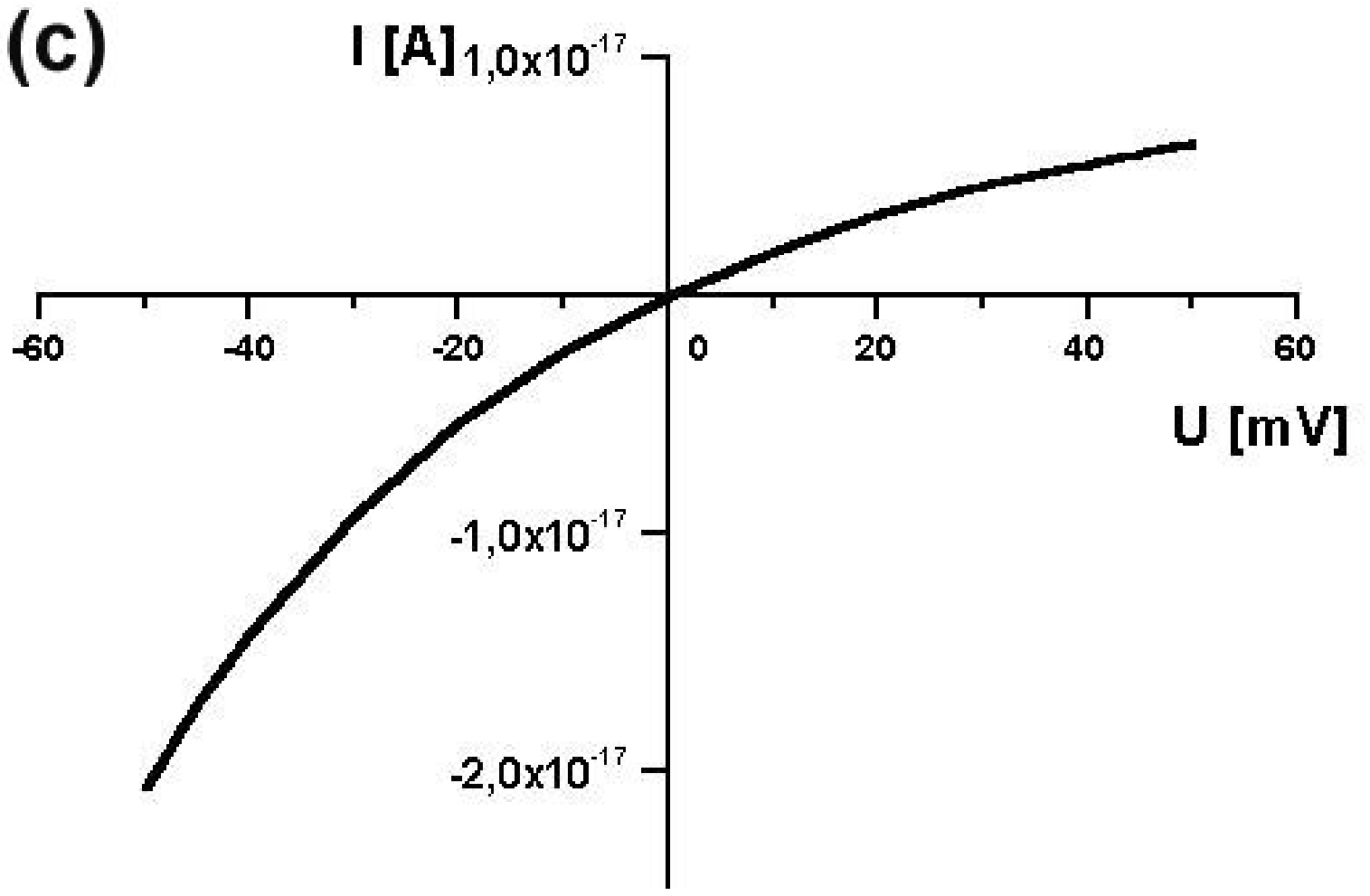}
      \includegraphics*[width=2.25in]{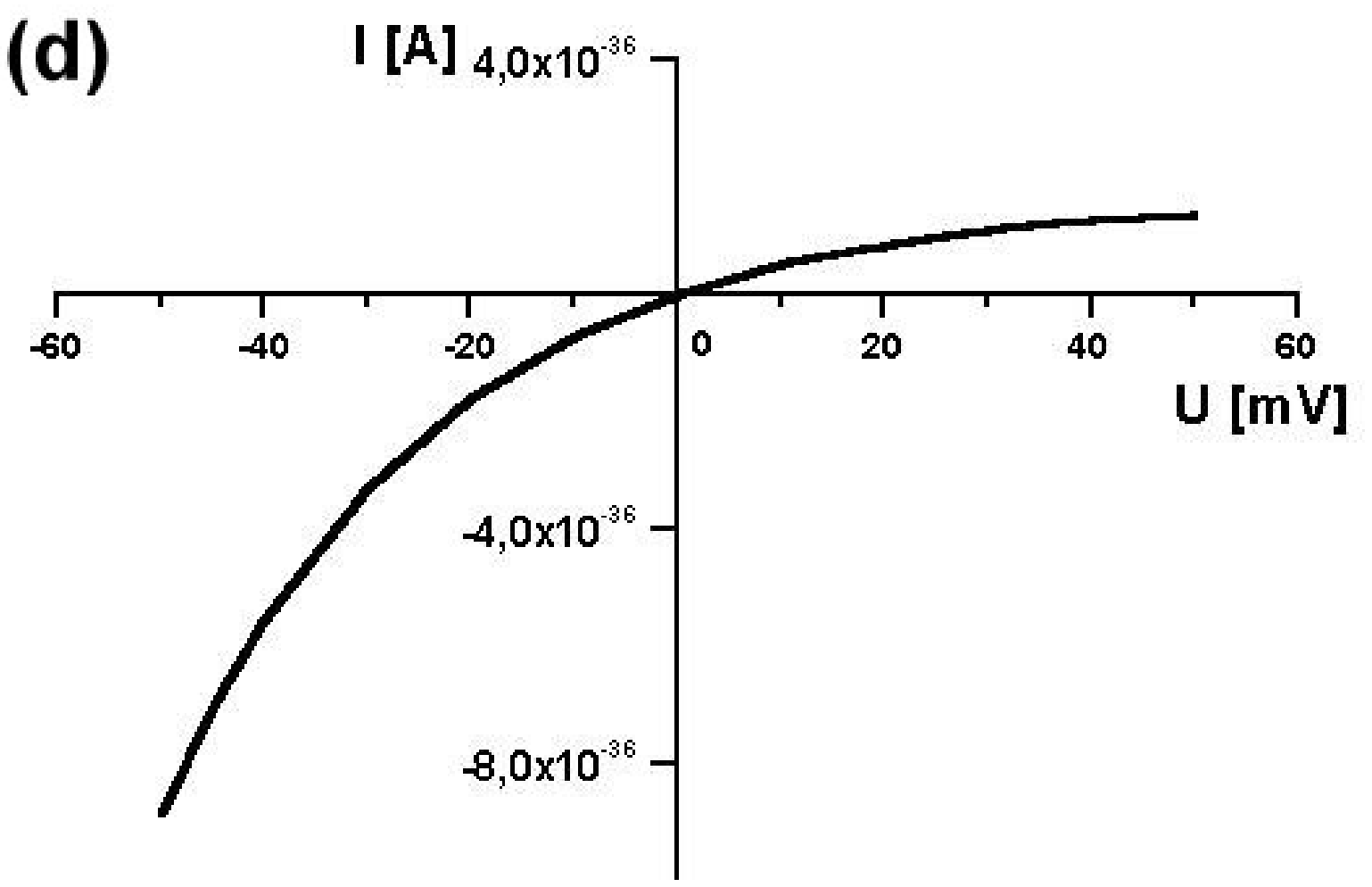}
       \caption{\small{$I-V$ dependence for the $\overline{\phi}_{int}$ being
ratchet-like function of $\overline{z}$, for
$\overline{\phi}_{int}(\overline{L}_1=0)=\overline{\phi}_{int}(\overline{L}_2=1)=0$,
$\overline{R}=R/L=0.0004$, $c_{0}=c_{1}=0.01$ M,
$D_{K^{+}}=2*10^{-9}$ m$^2$/s and $\overline{g}=0.00001$; a)
$\overline{\phi}(\overline{g})=-5$; b)
$\overline{\phi}(\overline{g})=-50$; c)
$\overline{\phi}(\overline{g})=5$; d)
$\overline{\phi}(\overline{g})=50$.}}
    \end{center}
    \label{fig:6}
\end{figure}

Finally, let us analyze the difference between the potential well
and the potential barrier inside the channel. The cation that
passes the negatively charged channel experiences potential well,
in the contrast the anion goes over the potential barrier. We can
see that the direction of the rectification is different in these
two cases (Fig. 6). What is more important, the current increases
with the growth of the depth of the potential well, whereas the
growth of the height of the potential barrier results in dramatic
decrease of the current (Fig. 6). The main point is that these two
cases (potential well and barrier) are not fully symmetric. The
contrast between these two cases leads us to the conclusion that
the same electric field could cause enhancement of the cation
current and inhibition of the anion current therefore the
resulting cation selectivity of the channel. However, it has been
shown recently that the selectivity of ionic channels may
originate from physical mechanisms different from the
electrostatics interactions, too \cite{Laio}.

\section{Comments and Conclusions}

We found that the asymmetry of the internal potential $\phi_{int}$
coupled with the potential well depth are sufficient for the
rectification observed as a diode-like shape of $I-V$
characteristic. The concentration gradient is not required for
that effect. However, in the absence of internal potential
$\phi_{int}(z)$), the Goldman-Hodgkin-Katz current equation,
Eq.~\ref{eqn:prad}, gives ohmic curve $I-V$ if $\Delta c=0$ and
non-ohmic shape if $\Delta c\ne 0$ \cite{Fall}. In the latter case
the deviation from ohmic behavior results from the gradient of
concentration on both sides of the channel and is weakly
non-linear. In our model, when we put $\Delta c=0$, the internal
potential $\overline{\phi}_{int}$ (with the boundary values
$\overline{\phi}_{int}(0)=\overline{\phi}_{int}(1)=0$) is the
source of asymmetry, and gives strong non-linearity of $I-V$. Note
that by varying these two factors the degree of the rectification
can be intensified or diminished.

The conclusions are supported by the fact that we solve the
Laplace's equation instead of the nonlinear Poisson-Boltzmann
equation (see Appendix). It allows us to avoid a misleading
intuition that a nonlinear diode-like shape of $I-V$
characteristics is caused by the non-linearity of
Poisson-Boltzmann equation. Apart from that, in general, it is not
possible to obtain an exact solution to the Nernst-Planck (NP)
equation, coupled to the electric field by means of the
Poisson-Boltzmann equation. The purpose of this paper is to show
that a simplified continuous model that makes use only of the
electrical properties of the open channel may support our
intuitive knowledge of examined phenomenon.

Moreover, our understanding of the mechanisms which account for
experimental observations will enable us to design both artificial
biological and synthetic channels with desired properties.
Engineered nanopores may have significant applications. The
properties of channels and pores that might be engineered include
conductance, ion selectivity, gating and rectification, and
inhibition by blockers \cite{Bayley}. This prospective
additionally confirms a need for the theoretical models of
nanopores.

\begin{acknowledgments}

The author would like to thank A. Fuli\'nski and M. Kotulska for
insightful comments and kind co-operation. Discussions with S.
Bezrukov, A. Berezhkovskii and G. Hummer are gratefully
acknowledged.

\end{acknowledgments}

\appendix*

\section{}

The cylinder has a radius $R$ and a height $L$, the top and bottom
surfaces being at $z=L$ and $z=0$. The surface charge density on
the side of the cylinder is $\sigma$ and we assume that is
symmetric about the axis of the cylinder, $\sigma=\sigma(z)$. We
want to find potential at any point inside the cylinder.

\begin{figure}[htb!]
    \begin{center}
      \includegraphics*[width=3.25in]{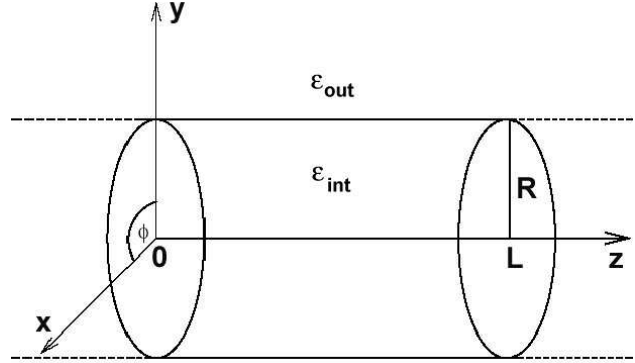}
       \caption{\small{Channel in cartesian coordinates ($x,y,z$) with the $z$
axis lying along the axis of the cylinder; the cylinder has a
radius $R$ and a length $L$; $\epsilon_{out}$, $\epsilon_{int}$
are dielectric constants outside and inside the channel,
respectively.}}
    \end{center}
    \label{fig:7}
\end{figure}

There is no loss of generality if we take the infinitely long
cylinder charged to prescribed charge density $\sigma$, which can
differ from 0 only for $z\in [0,L]$ (Fig. 7). In cylindrical
coordinates $(r, \theta, z)$ the Laplace's equation takes form
\begin{equation}
\left[\frac{\partial^2}{\partial
r^2}+\frac{1}{r}\frac{\partial}{\partial
r}+\frac{1}{r^2}\frac{\partial^2}{\partial
\theta^2}+\frac{\partial}{\partial z^2}\right]\phi=0.
\end{equation}
The separation of variables is accomplished by the substitution
$\phi (r,z)=P(r)Z(z)$. This leads to the two ordinary differential
equations:
\begin{equation}
\begin{array}{c}
\vspace{10pt} \displaystyle\frac{\partial^2 P}{\partial
r^2}+\frac{1}{r}\frac{\partial
P}{\partial r}-k^2 P=0\\
\displaystyle\frac{\partial Z}{\partial z^2}+k^2Z=0.
\end{array}
\end{equation}
We denote by $\phi_{int}$ the potential inside the channel and by
$\phi_{out}$ the potential outside the channel:
\begin{equation}
\begin{array}{c}
\vspace{5pt} \phi_{int}(r,z) = \int_{-\infty}^{+\infty} dk
[A(k)e^{ikz}+B(k)e^{-ikz}]I_0(|k|r)\\
\phi_{out}(r,z) = \int_{-\infty}^{+\infty} dk
[C(k)e^{ikz}+D(k)e^{-ikz}]K_0(|k|r).
\end{array}
\end{equation}
Further, these functions must satisfy the boundary conditions
\begin{equation}
\begin{array}{c}
\vspace{10pt}
\phi_{int}(r,z) = \phi_{out}(r,z)~~~~~\rm{on}~~r=R\\
\displaystyle\epsilon_{out}\frac{\partial\phi_{out}(r,z)}{\partial
r} -\epsilon_{int}\frac{\partial\phi_{int}(r,z)}{\partial
r}=-\sigma(z)~~~~~\rm{on}~~r=R,
\end{array}
\end{equation}
where $\epsilon_{out},~\epsilon_{int}$ are the dielectric
constants outside and inside the channel, respectively.

The result on the $z$ axis is
\begin{equation}
\phi^{r=0}_{int}(z) =
-\frac{1}{\sqrt{2\pi}}\int_{-\infty}^{+\infty} dk
e^{ikz}\hat{\sigma}(k)\mathcal{W}(|k|),\label{eqn:app}
\end{equation}
where
\begin{equation}
\mathcal{W}(|k|)=\frac{K_0(|k|R)}{|k|
[\epsilon_{int}I_1(|k|R)K_0(|k|R)+\epsilon_{out}I_0(|k|R)K_1(|k|R)]}
\end{equation}
and
\begin{equation}
\hat{\sigma}(k) = \frac{1}{\sqrt{2\pi}}\int_{-\infty}^{+\infty} dz
\sigma(z) e^{-ikz}.\label{eqn:pow}
\end{equation}


\begin{thebibliography}{99}
\bibitem{Apel}
P.Yu. Apel, Yu.E. Korchev, Z. Siwy, R. Spohr, M. Yoshida, Nucl.
Instrum. Meth. {\bf B 184}, 337 (2001).

\bibitem{Siwy}
Z. Siwy, Y. Gu, H. Spohr, D. Baur, A. Wolf-Reber, R. Spohr, P.
Apel, Y.E. Korchev, Europhys. Lett. {\bf 60}, 349 (2002).

\bibitem{Eyring}
B.J. Zwolinski and H.Eyring and C. E. Reese, J. Phys. Colloid
Chem., {\bf 53}, 1426-1453 (1949).

\bibitem{Lauger}
P. Läuger, Biochim. Biophys. Acta., {\bf 311}, 423-441 (1972).

\bibitem{Woodbury}
J. W. Woodbury, {\it Eyring-rate theory model of the
current-voltage relationships of ion channels in excitable
 membranes}, in Chemical Dynamics: Papers in Honor of
Henry Eyring, edited by J. O. Hirschfelder (Wiley, New York, 1971)

\bibitem{Sneyd}
J. Keener, J. Sneyd {\it Mathematical Physiology} (Springer,
Sunderland, MA, 1992), 2nd ed.

\bibitem{Nonner}
W. Nonner, D. P. Chen, and B. Eisenberg, J. Gen. Physiol. {\bf
113}, 773-782 (1999).

\bibitem{Corry}
B.Corry, S.Kuyucak, and S.-H. Chung, Biophys. J., {\bf 78}, 2364
(2000).

\bibitem{Corry2}
B.Corry and S.Kuyucak and S.-H. Chung, J. Gen. Physiol., {\bf
114}, 597-599 (1999).

\bibitem{Lewitt}
D. G. Lewitt, J. Gen. Physiol. {\bf 113}, 789 (1999).

\bibitem{Laio}
A. Laio and V. Torre, Biophys. J., {\bf 76}, 129-148 (1999).

\bibitem{Siwy2}
Z. Siwy, P. Apel, D. Baur, D. D. Dobrev, Y. E. Korchev, R.
Neumann, R. Spohr, C. Trautmann, K. Voss, Surface Science {\bf
532-535}, 1061 (2003).

\bibitem{Siwy4}
Z. Siwy, A. Fuli\'nski, Phys. Rev. Lett. {\bf 89}, 158101 (2002).

\bibitem{kienker}
P.K. Kienker, W.F. DeGrado, and J.D. Lear, Proc. Natl. Acad. Sci.
USA {\bf 91}, 4859 (1994).

\bibitem{okazaki}
T. Okazaki, M. Sakoh, Y. Nagaoka, and K. Asami, Biophys. J. {\bf
85}, 267 (2003).

\bibitem{Gor}
E.Gorczy\'nska, P.L. Huddie, B.A. Miller, I.R. Mellor, R.L.
Ramsey, and P.N.R. Usherwood, Pfl\"ugers Arch. Ges. Physiol.
Menschen Tiere, {\bf 432}, 597 (1996).

\bibitem{Siwy3}
Z. Siwy, D.D. Dobrev, R. Neumann, C. Trautmann, K. Voss, Applied
Physics A {\bf 76}, 781 (2003).

\bibitem{Ful1}
A. Fuli\'nski, I. D. Kosi\'nska, and Z. Siwy, Europhys. Lett.
{\bf67}, 683 (2004).

\bibitem{SiF}
Z. Siwy, A. Fuli\'nski, Phys. Rev. Lett. {\bf 89}, 198103 (2002).

\bibitem{Kos1}
Z. Siwy, I.D. Kosi\'nska, A. Fuli\'nski, and C.R.Martin, Phys.
Rev. Lett. {\bf 94}, 048102 (2005).

\bibitem{Kos2}
I.D. Kosi\'nska, A. Fuli\'nski, Phys. Rev. E {\bf 72}, 011201
(2005).

\bibitem{Hanggi}
P. H\"anggi and R. Bartussek, {\it Brownian rectifiers: How to
convert Brownian motion into directed transport}, in Nonlinear
Physics of Complex Systems, edited by J. Parisi, S. C. M\"uller,
and W. Zimmermann (Springer, Berlin, 1997), Vol. 476, pp. 294-308;
R. Bartussek, P. H\"anggi, and J.G. Kissner, Europhys. Lett. {\bf
29}, 459 (1994).

\bibitem{Marquet}
C. Marquet, A. Buguin, L. Talini, and P. Silberzan, Phys. Rev.
Lett. {\bf 88}, 168301 (2002).

\bibitem{Mon}
K.K. Mon, J.K. Percus, J.Chem. Phys. {\bf 122}, 214503 (2005).

\bibitem{Ful}
A. Fuli\'nski, Acta Phys. Pol. {\bf 29}, 1523 (1998).

\bibitem{Smol}
M. Smoluchowski, Ann. Physik {\bf48}, 1103 (1915); Phys. Z.
{\bf17}, 557, 585 (1916).

\bibitem{Smol2}
M. Smoluchowski, Phys. Z., {\bf 17}, 557-585 (1916).

\bibitem{Kuyucak}
S. Kuyucak, O. S. Andersen, and S-H. Chung, Rep. Prog. Phys.
{\bf64}, 1427 (2001).

\bibitem{1}
D. Appel, Nature {\bf 419}, 1300 (1991); E. Toimil-Molares et al.,
Adv. Mater. {\bf 13}, 62 (2001); E. Toimil-Molares et al., Nucl.
Instr. and Meth. B {\bf 185}, 192 (2001).

\bibitem{NJP}
A. Fuli\'nski, I.  Kosi\'nska, and Z. Siwy, New J. Phys. {\bf 7},
132 (2005).

\bibitem{Schattat}
B. Schattat, W. Bolse, S. Klaumunzer, I. Zizak, R. Scholz, Appl.
Phys. Lett. {\bf 87}, 173110 (2005).

\bibitem{Woolf}
T.B. Woolf and B. Roux, Proteins Struct. Funct. Genet. {\bf 24},
92 (1996).

\bibitem{porin}
J. Song, C.A.S.A. Minetti, M.S. Blake and M. Colombini, Biophys.
J. {\bf 76}, 804 (1999).

\bibitem{Hille}
B. Hille, {\it Ionic Channels of Excitable Membranes} (Sinauer,
Sunderland, MA, 1992), 2nd ed.

\bibitem{Lev}
A.A. Lev, Y.E. Korchev, T.K. Rostovtseva, C.L. Bashford, D.T.
Edmonds, and C.A. Pasternak, Proc. R. Soc. Lond. B {\bf 252}, 187
(1993).

\bibitem{Cole}
K. Cole, {\it A Quantitative Description of Membrane Current and
Its Application to Conductance and Excitation in Nerve.}
(University of California Press, Berkeley 1968).

\bibitem{Fall}
Ch.P. Fall, E.S. Marland, J.M. Wagner, J.J. Tyson, {\it
Computational Cell Biology} (Springer-Verlag New York, Inc. 2002).

\bibitem{Bayley}
H. Bayley and L. Jayasinghe, Mol. Membrane Biol., {\bf 21},
209-220 (2004).

\end{thebibliography}
\end{document}